\documentclass[onecolumn]{aastex631}

\usepackage{epsfig}
\usepackage{xcolor}
\usepackage{amssymb}
\usepackage{pifont}

\usepackage{romannum}
\usepackage{lipsum}
\usepackage{amsmath}

\received{}
\revised{}
\accepted{}
\submitjournal{ApJ}

\shorttitle{Optimal Strategies for a Pulsar Timing Array}
\shortauthors{T.~Liu et al.}

\begin{document}

\title{ Multimessenger Approaches to Supermassive Black Hole Binary Detection and Parameter Estimation II: Optimal Strategies for a Pulsar Timing Array }

\correspondingauthor{Tingting Liu}
\email{tingting.liu@nanograv.org}

\author[0000-0001-5766-4287]{Tingting Liu}
\affiliation{Center for Gravitation, Cosmology and Astrophysics, Department of Physics, University of Wisconsin-Milwaukee, Milwaukee, WI 53211, USA}
\affiliation{Department of Physics and Astronomy, West Virginia University, P.O. Box 6315, Morgantown, WV 26506, USA}

\author{Tyler Cohen}
\affiliation{Department of Physics, New Mexico Institute of Mining and Technology, 801 Leroy Place, Socorro, NM 87801, USA}

\author[0000-0002-6155-3501]{Casey McGrath}
\affiliation{Center for Space Sciences and Technology, University of Maryland, Baltimore County, Baltimore, MD 21250, USA}
\affiliation{Gravitational Astrophysics Lab, NASA Goddard Space Flight Center, Greenbelt, MD 20771, USA}
\affiliation{Center for Research and Exploration in Space Science and Technology II, NASA Goddard Space Flight Center, Greenbelt, MD 20771, USA}

\author{Paul B. Demorest}
\affiliation{National Radio Astronomy Observatory, 1003 Lopezville Rd., Socorro, NM 87801, USA}

\author{Sarah Vigeland}
\affiliation{Center for Gravitation, Cosmology and Astrophysics, Department of Physics, University of Wisconsin-Milwaukee, Milwaukee, WI 53211, USA}


\begin{abstract}
Pulsar timing arrays (PTAs) are Galactic-scale gravitational wave (GW) detectors consisting of precisely-timed pulsars distributed across the sky. Within the decade, PTAs are expected to detect the nanohertz GWs emitted by close-separation supermassive black hole binaries (SMBHBs), thereby opening up the low frequency end of the GW spectrum for science. Individual SMBHBs which power active galactic nuclei are also promising multi-messenger sources; they may be identified via theoretically predicted electromagnetic (EM) signatures and be followed up by PTAs for GW observations. In this work, we study the detection and parameter estimation prospects of a PTA which targets EM-selected SMBHBs. Adopting a simulated Galactic millisecond pulsar population, we envisage three different pulsar timing campaigns which observe three mock sources at different sky locations. We find that an all-sky PTA which times the best pulsars is an optimal and feasible approach to observe EM-selected SMBHBs and measure their source parameters to high precision (i.e., comparable to or better than conventional EM measurements). We discuss the implications of our findings in the context of the future PTA experiment with the planned Deep Synoptic Array-2000 and the multi-messenger studies of SMBHBs such as the well-known binary candidate OJ 287.
\end{abstract}


\section{Introduction} \label{sec:intro}

The detection and characterization of low-frequency GWs are considered the next frontiers of GW astronomy. This frequency range can be probed by two GW experiments:  the Laser Interferometer Space Antenna (\citealt{Amaro-Seoane2017}), which is expected to launch in the mid-2030s and will probe GW sources emitting at mHz frequencies; and pulsar timing arrays (PTAs; e.g., \citealt{NANOGrav,EPTA,PPTA}), which have been operating since the 2000s and are expected to reach the sensitivity for nanohertz GW science in this decade. 

The main astrophysical sources which emit GWs in the nanohertz range ($\sim 10^{-9}-10^{-7}$ Hz) are supermassive black hole binaries (SMBHBs) which are thought to form following galaxy mergers. As a population, their demographics and properties encode information about, e.g., the relationship between galaxies and their central SMBHs and the timescale of SMBH mergers. As individual sources, SMBHBs may power active galactic nuclei (AGN) through accretion and serve as interesting and rare astrophysical laboratories to study the accretion of matter onto binary black holes. Therefore, the GW observations of individual SMBHBs will also open up new areas in multi-messenger astrophysics. 

While it is anticipated that PTAs may soon detect the gravitational wave background (GWB), which is the ensemble signal from a cosmological population of SMBHBs (e.g., \citealt{Taylor2016,Kelley2017GWB,Pol2021}), the best path toward detecting continuous wave (CW) signals from individual sources is less clear. By nature, CW sources are highly anisotropic, which could require different observing strategies than simply timing a large number of pulsars distributed across the sky and adding more to the array over time (which have been the main strategies toward the observations of an isotropic GWB). For example, should we preferentially add to the PTA pulsars near a possible SMBHB? Should the limited amount of total observing time be allocated to focus on timing a small number of pulsars with the lowest timing noise? Or is an all-sky PTA which times as many pulsars as possible the best (and a feasible) approach?

Those possible strategies have important implications for the multi-messenger studies of SMBHBs. Many binary candidates whose orbital periods would be $\sim$ months to years (i.e., corresponding to the PTA GW frequency range) have so far been reported in the literature (e.g., \citealt{Graham2015Nat,Graham2015,Liu2015,Liu2016,Liu2019,Charisi2016,Severgnini2018}). Many of those candidates were identified thanks to modern synoptic surveys which monitor a large number of AGN, which makes it possible to systematically search for the intrinsically rare SMBHBs. Additionally, the typical cadence and length of a time-domain survey make it sensitive to periodic variability on $\sim$ year timescales arising from mechanisms such binary-modulated accretion (e.g., \citealt{Noble2012,D'Orazio2013,Farris2014}), Doppler beaming \citep{D'Orazio2015}, and self-lensing \citep{D'Orazio2018selflensing}. The upcoming Rubin Observatory Legacy Survey of Space and Time (LSST; \citealt{Ivezic2008}) could further increase the number of known binary candidates manyfold \citep{Kelley2018SMBHB,Kelley2021}. Unfortunately, the EM follow-up or confirmation of current candidates have been less than successful (e.g., \citealt{Foord2017,Saade2020,Guo2020}) due to the large false positive rate in systematic searches, the lack of complementary evidence in support of the binary hypothesis, and the difficulty of distinguishing the EM emission of an SMBHB from that of a single AGN. The GW observations of EM-selected SMBHB candidates would therefore lead to at least three important breakthroughs: (1) directly and unequivocally confirming the nature of the sources via the detection of GWs, (2) studying their properties through parameter measurements via GWs and at the same time independently verifying the source parameters obtained from conventional EM observations, and (3) breaking degeneracies and extracting astrophysical information through combined GW and EM observations and connecting observational properties to the theoretical models of accreting SMBHBs. Furthermore, both detection signal-to-noise (SNR) and parameter measurement uncertainties can be improved by searching for GWs at the sky location of the EM counterpart \citep{Liu2021}, thereby allowing the studies of low-SNR sources which would otherwise be missed in unguided all-sky searches.

While none of the reported binary candidates are within the current PTA sensitivity, it is nevertheless important to understand the capability of a given PTA experimental setup in order to understand what astrophysical information we will be able to extract from it. For example, \cite{Sesana2010} have studied the binary parameter measurement uncertainties as a function of, e.g., number of pulsars in the array and their sky coverage; they have also investigated the sky localization capability which would be important for EM followup. Additionally, we can apply the knowledge of PTA capabilities to either fine-tune the current PTAs in operation, or design the next-generation PTA experiment. For instance, \cite{Burt2011} and \cite{Christy2014} have found that increasing observing time on the best-timed pulsars would have the best effect on the PTA sensitivity volume. Based on this suggestion, the North American Nanohertz Observatory for Gravitational Waves (NANOGrav) started a high-cadence timing campaign of six pulsars beginning in 2013 (at the Green Bank Telescope) and 2015 (at the Arecibo Telescope) to increase its sensitivity to CWs\footnote{However, the program was suspended in 2020 due to the collapse of Arecibo.} \citep{NG11yrdata}.

In this work, we aim to study the effect of different PTA configurations on the measurement of binary parameters and explore strategies to optimize the PTA experiment for the observations of SMBHBs. Our study differs from previous work in several ways: first, we assume we have already identified an EM candidate and the PTA data are being searched to study the ``GW counterpart'' of the source. Second, we explore both detection and parameter estimation aspects of the optimized PTA. Finally, our mock PTAs are based on realistic pulsar populations and properties, as well as observational considerations. This paper is organized as follows: in \S \ref{sec:methods}, we describe our methodology of constructing a simulated PTA and measuring binary parameters; in \S \ref{sec:results}, we present the results of observing three mock SMBHB sources with different PTA setups; we summarize in \S \ref{sec:conclude} and make suggestions for PTA designs which optimize single source multi-messenger observations. Throughout the paper, we adopt geometrized units: G=c=1.


\section{Methods} \label{sec:methods}

\subsection{Constructing a mock PTA dataset using a simulated pulsar population} \label{sec:pta}

A PTA regularly monitors a number of millisecond pulsars (MSPs), which are characterized by rapid and extremely stable rotation periods. It is therefore sensitive to fluctuations in the arrival times of the radio pulses induced by GWs emitted by astrophysical sources including SMBHBs. We first simulate ten realizations of the MSP population using the pulsar population synthesis package \texttt{PsrPopPy}\footnote{https://github.com/samb8s/PsrPopPy} \citep{Bates2014} in the snapshot mode, where simulated MSPs are generated by randomly drawing properties from distributions constrained by earlier studies. The overall population size is constrained by halting generation when the population contains the same number of MSPs (28) that would have been discovered by the Parkes Multibeam Survey \citep{ParkesMBI}. We assume MSPs are distributed radially from the Galactic Center according to a Gaussian distribution with a standard deviation of 7.5 kpc and exponentially above and below the galactic plane with a mean scale height of 500 pc \citep{lorimer12_IAU}. We assume that MSP spin periods are distributed according to \citet{lorimer12_IAU} and that their luminosities (at 1.4 GHz) are distributed according to a log-normal distribution with $\langle \log_{10} \rm L\rangle = -1.1$ and $\sigma_{\log_{10} \rm L} = 0.9$ \citep{fk06}, where $L$ is in units of $\rm mJy\ \rm kpc^2$. \citet{aggarwal22} found that MSP spectral indices are best described by a broken power-law with a break at 300 MHz, but since all of our simulated observations occur above 300 MHz, we simply use the high frequency part of their model: Gaussian with mean spectral index $-1.47$ and standard deviation $0.7$. We assume that MSP pulse widths are distributed $\mathrm{log_{10}}$-normally with a mean of $0.05P^{0.9}$ and a standard deviation of 0.3. Finally, dispersion measures (DM) and scattering timescales are computed by the NE2001 electron density model of the Milky Way \citep{ne2001}.

\begin{figure*}[ht]
\centering
\epsfig{file=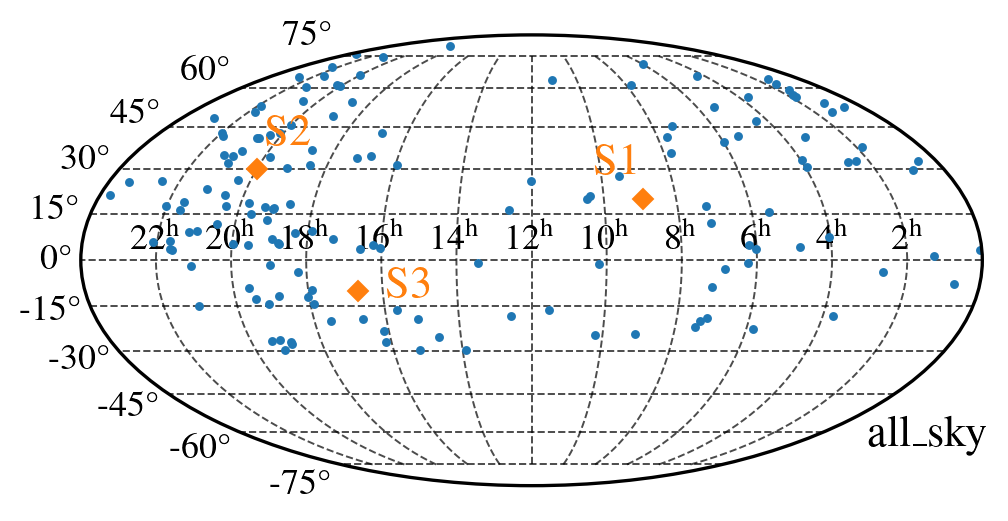,width=0.4\textwidth,clip=}
\epsfig{file=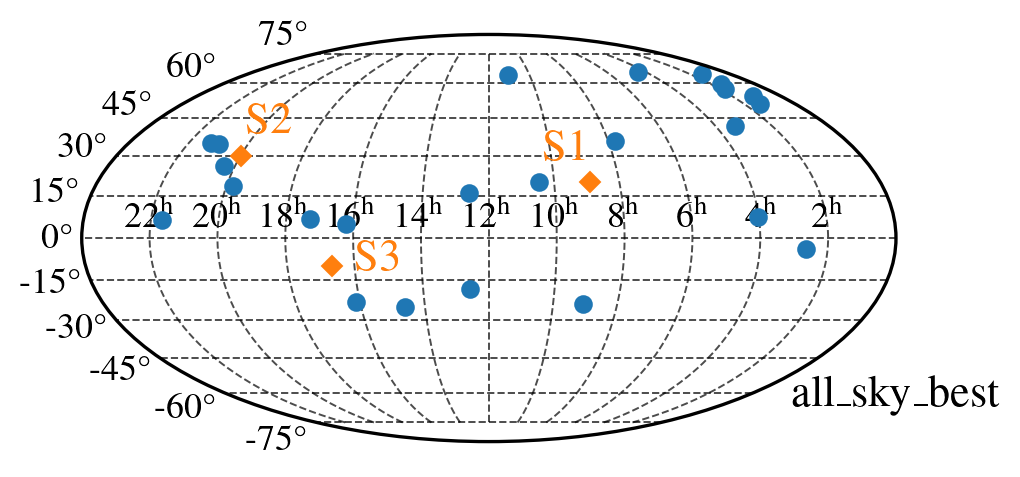,width=0.45\textwidth,clip=}
\epsfig{file=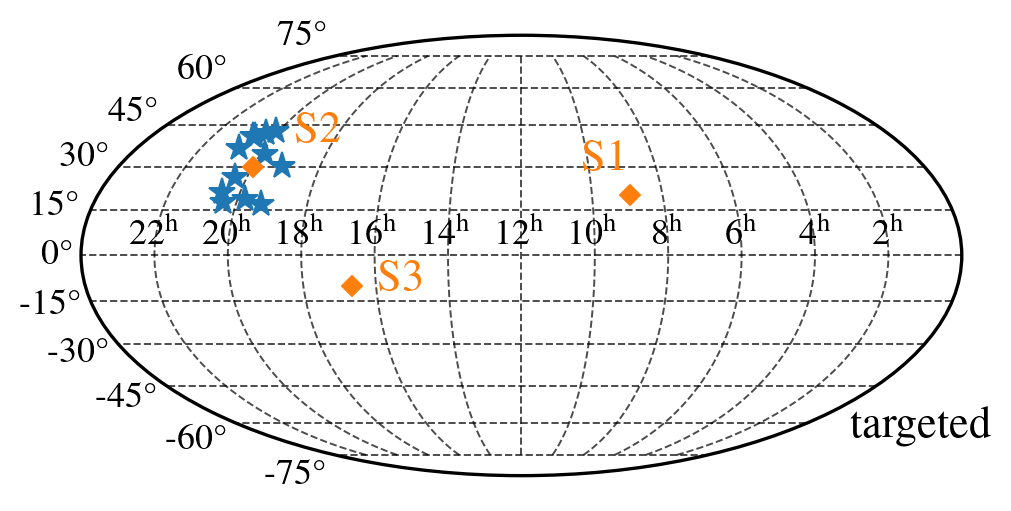,width=0.45\textwidth,clip=}
\caption{In the first panel, we show one realization of the simulated MSP population with $\sigma_{\rm TOA}<1 \mu$s observed by DSA-2000 (small blue filled circles); those MSPs form the \texttt{all\_sky} PTA. In the second panel, we show the ones with $\sigma_{\rm TOA}<100$ ns as large filled circles (\texttt{all\_sky\_best}). In the last panel, we show the ones which are located near S2 as stars (\texttt{targeted}). The three sources whose detection and parameter estimation prospects are examined in this work are shown as orange diamonds.}
\label{fig:map}
\end{figure*}

\begin{figure*}[ht]
\centering
\epsfig{file=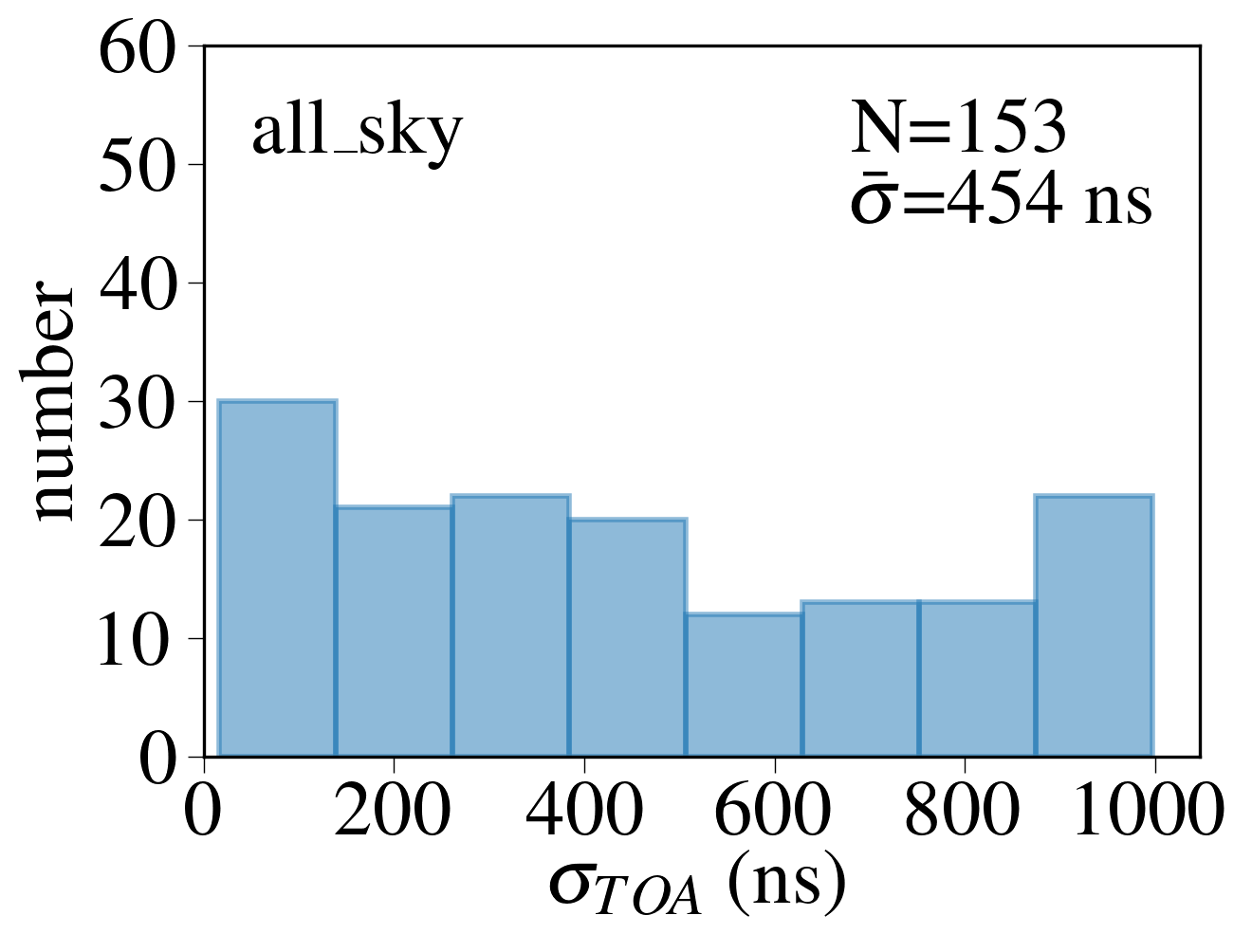,width=0.32\textwidth,clip=}
\epsfig{file=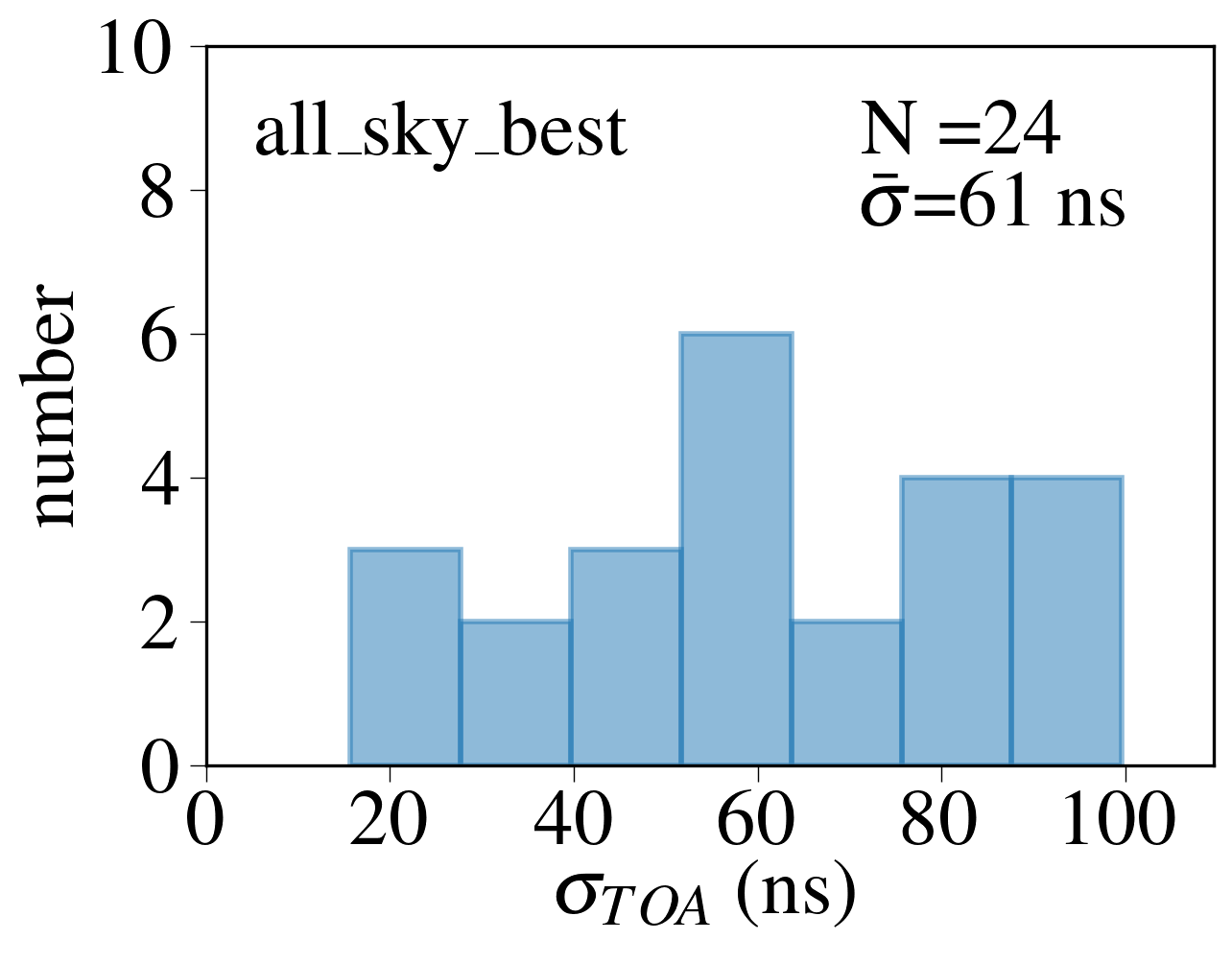,width=0.32\textwidth,clip=}
\epsfig{file=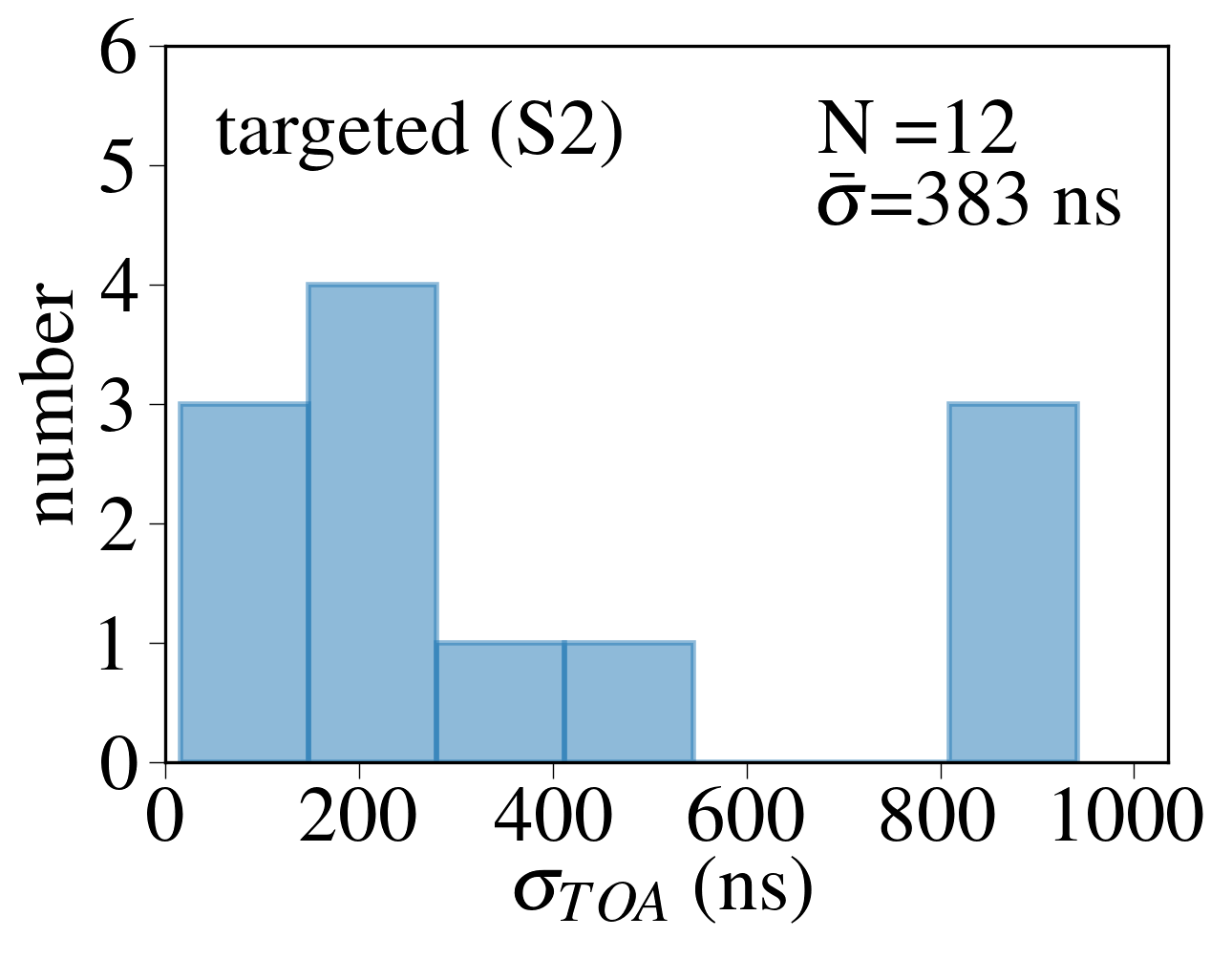,width=0.32\textwidth,clip=}
\caption{We show the respective $\sigma_{\rm TOA}$ distributions, sample sizes, and mean $\sigma_{\rm TOA}$ of the three pulsar samples shown in Figure \ref{fig:map}.}
\label{fig:sigma_hist}
\end{figure*}

In order to construct a PTA from the simulated MSP population, we simulate a timing program with the proposed Deep Synoptic Array-2000 (DSA-2000; \citealt{Hallinan2019}), where an anticipated $25\%$ of the time will be used for pulsar timing. We simulate timing observations of each MSP using the full DSA-2000 collecting area operating in the frequency range $0.7-2$ GHz and assume an integration time of 60 minutes per pulsar per observation epoch. We assume that DSA-2000 will have a system temperature of $25\ \mathrm{K}$, a gain of $10\ \mathrm{K\ Jy^{-1}}$, and will be able to observe declinations from $-30^{\circ}$ to $90^{\circ}$. If a model pulsar is within the declination limits, its pulse duty cycle is $<90\%$, and it is sufficiently bright to be detected, we then include it in our mock PTA and compute the uncertainty on its pulse time of arrival $\sigma_\mathrm{TOA}$ which is dependent on both pulsar and telescope parameters. To do so, we use the FrequencyOptimizer\footnote{https://github.com/mtlam/FrequencyOptimizer} software package \citep{lam18_optimum_freq}, which defines the TOA uncertainty as\footnote{The total $\sigma_{\rm TOA}$ in \citet{lam18_optimum_freq} originally contains an additional term, $\sigma_\mathrm{W}$, containing sources of white noise. We ultimately determined that since $\sigma_{\widehat{DM}}$ is derived from the white noise covariance matrix, adding $\sigma_\mathrm{W}$ in quadrature would double-count these contributions so we omit that term in our analysis.}
\[
\sigma_\mathrm{TOA}^2 = \sigma_{\widehat{DM}}^2 + \sigma_{\delta t_C}^2 + 
	\sigma_{\mathrm{DM}(\nu)}^2 + \sigma_\mathrm{tel}^2 \, .
\]
$\sigma_{\widehat{DM}}$ is the standard error on the infinite-frequency TOA when fitting the timing residuals for DM; it contains sources of white noise including template-fitting error, scintillation error, and jitter error, where the jitter error is computed using the scaling relationship shown in Figure 7 of \cite{lam19_jitter},
\[
\sigma_J = 10^{-0.218}\delta^{1.216}\frac{P^{3/2}}{\sqrt{t_\mathrm{int}}} \, ,
\]

\noindent which is a function of pulsar spin period $P$, pulse duty cycle $\delta$, and integration time $t_\mathrm{int}$. $\sigma_{\delta t_C}$ is the systematic error due to chromatic variations in pulse width caused by interstellar scattering and dispersion. $\sigma_{\mathrm{DM}(\nu)}$ is the uncertainty in DM estimation between two frequency channels. $\sigma_\mathrm{tel}$ contains uncertainties due to radio frequency interference, incorrect gain calibration, and instrumental self-polarization. NANOGrav considers $\sigma_{\rm TOA}<1 \mu s$ to be the minimum criterion for inclusion of an MSP in the PTA, so we adopt that cutoff here. The result are ten mock PTAs with an average of $\sim 180$ MSPs with $\sigma_{\rm TOA}<1 \mu s$. The sky distribution and distribution of $\sigma_{\rm TOA}$ for one realization are plotted in the first panel of Figure \ref{fig:map} and the left panel of Figure \ref{fig:sigma_hist} (\texttt{all\_sky}) respectively.

    \startlongtable
    \begin{splitdeluxetable}{ccccccccccccBcccccccccccc}
        \tabletypesize{\scriptsize}
        \tablecaption{Sample of Simulated Pulsars}
        \tablehead{
            $P$ &	$\dot{P}$	&	DM	&	$t_\mathrm{scatter}$	 &	$W_{50}$ 	&	$l$	&	$b$ 	&	RA	 &	Dec.  &	$S_{1400}$ &	$L_{1400}$ 	&	$\alpha$	&	$d_\mathrm{true}$ 	&	X 	&	Y 	&	Z 	&	$\sigma_\mathrm{TOA}$ 	&	$\sigma_\mathrm{tel}$ 	&	$\sigma_\mathrm{\delta DM}$	&	$\sigma_\mathrm{W}$ 	&	$\sigma_\mathrm{J}$ 	&	$A_\mathrm{red}$ 	&	$\gamma_\mathrm{red}$	&	$A_\mathrm{eff} / A_\mathrm{eff, tot}$ \\
            (ms)	&	($\mathrm{s\ s^{-1}}$)	&	($\mathrm{pc\ cm^{-3}}$)	&	(ms)	&	(ms)	&	(deg)	&	(deg)	&	(deg)	&	(deg)	&	(mJy)	&	($\mathrm{mJy\ kpc^2}$)	&		&	(kpc)	&	(kpc)	&	(kpc)	&	(kpc)	&	($\mathrm{\mu s}$)	&	($\mathrm{\mu s}$)	&	($\mathrm{\mu s}$)	&	($\mathrm{\mu s}$)	&	($\mathrm{\mu s}$)	&	($\mathrm{\mu s\ yr^{1/2}}$)	&		&	\\
           }
           \startdata
6.64	&	2.19E-21	&	615.09	&	7.42	&	0.80	&	-0.39	&	1.85	&	264.39	&	-28.29	&	9.11E-04	&	0.08	&	-0.91	&	9.18	&	-0.06	&	-0.67	&	0.30	&	-1.00	&	-1.00	&	-1.00	&	-1.00	&	1.09	&	6.50E-04	&	-5.60	&	1.00\\ 
3.19	&	1.86E-19	&	177.42	&	2.01	&	0.06	&	6.82	&	-9.80	&	279.87	&	-27.63	&	0.004	&	0.15	&	-2.08	&	5.86	&	0.69	&	2.77	&	-1.00	&	1757.16	&	0.01	&	1757.16	&	5.68	&	0.05	&	0.09	&	-5.60	&	1.00\\ 
3.27	&	9.73E-21	&	336.38	&	3.86	&	0.11	&	-15.20	&	-7.19	&	264.06	&	-45.61	&	0.002	&	0.33	&	-2.45	&	14.57	&	-3.79	&	-5.45	&	-1.82	&	-2.00	&	-2.00	&	-2.00	&	-2.00	&	0.11	&	0.02	&	-5.60	&	1.00\\ 
4.42	&	1.37E-19	&	112.95	&	0.00	&	0.48	&	33.95	&	15.35	&	269.57	&	7.82	&	8.00E-04	&	0.03	&	-2.70	&	6.26	&	3.37	&	3.49	&	1.66	&	-1.00	&	-1.00	&	-1.00	&	-1.00	&	0.53	&	0.02	&	-5.60	&	1.00\\ 
14.04	&	4.74E-21	&	638.35	&	10.42	&	0.50	&	-3.47	&	0.52	&	263.76	&	-31.60	&	0.012	&	0.71	&	-3.04	&	7.63	&	-0.46	&	0.89	&	0.07	&	-2.00	&	-2.00	&	-2.00	&	-2.00	&	1.00	&	8.17E-04	&	-5.60	&	1.00\\ 
25.52	&	7.16E-20	&	1415.46	&	1143.50	&	0.79	&	12.99	&	-0.46	&	273.89	&	-17.89	&	2.26E-04	&	0.08	&	-0.18	&	19.35	&	4.35	&	-10.36	&	-0.16	&	-1.00	&	-1.00	&	-1.00	&	-1.00	&	2.12	&	1.95E-04	&	-5.60	&	1.00\\ 
7.00	&	7.94E-21	&	552.88	&	1514.56	&	0.21	&	54.88	&	0.25	&	293.04	&	19.55	&	2.54E-06	&	7.14E-04	&	-2.13	&	16.76	&	13.71	&	-1.14	&	0.07	&	-1.00	&	-1.00	&	-1.00	&	-1.00	&	0.29	&	0.00	&	-5.60	&	1.00\\ 
14.44	&	2.79E-19	&	943.11	&	500.37	&	0.29	&	-43.43	&	-0.23	&	221.07	&	-60.08	&	0.002	&	0.54	&	-1.69	&	15.14	&	-10.41	&	-2.50	&	-0.06	&	-2.00	&	-2.00	&	-2.00	&	-2.00	&	0.58	&	0.02	&	-5.60	&	1.00\\ 
28.56	&	5.91E-21	&	215.86	&	0.04	&	1.29	&	4.94	&	10.38	&	259.74	&	-19.19	&	0.006	&	0.52	&	-0.80	&	9.23	&	0.78	&	-0.54	&	1.66	&	101.11	&	0.04	&	101.11	&	27.34	&	3.62	&	3.04E-04	&	-5.60	&	1.00\\ 
1.98	&	6.71E-21	&	372.40	&	1.68	&	0.34	&	-2.35	&	4.21	&	260.91	&	-28.64	&	0.002	&	0.16	&	-0.64	&	8.46	&	-0.35	&	0.07	&	0.62	&	-1.00	&	-1.00	&	-1.00	&	-1.00	&	0.25	&	0.01	&	-5.60	&	1.00\\ 
 \enddata
 	\tablenotetext{}{$P$ -- pulse period}
	\tablenotetext{}{$\dot{P}$ -- period derivative}
	\tablenotetext{}{DM -- NE2001 dispersion measure}
	\tablenotetext{}{$t_\mathrm{scatter}$-- scattering timescale}
	\tablenotetext{}{$W_{50}$ -- intrinsic pulse full-width at half-maximum}
	\tablenotetext{}{$l$, $b$ -- galactic longitude and latitude. Note that in \texttt{PsrPopPy}, $-180^{\circ}<l<+180^{\circ}$.}
	\tablenotetext{}{RA, Dec -- right ascension and declination}
	\tablenotetext{}{$S_{1400}$ -- flux at 1.4 GHz}
	\tablenotetext{}{$L_{1400}$ -- luminosity at 1.4 GHz}
	\tablenotetext{}{$\alpha$ -- spectral index}
	\tablenotetext{}{$d_\mathrm{true}$ -- true distance from Earth}
	\tablenotetext{}{X, Y, Z -- galactic cartesian positions. The Sun is at X=0, Y=8.5 kpc, Z=0 in this coordinate system.}
	\tablenotetext{}{$\sigma_\mathrm{TOA}$ -- total TOA uncertainty}
	\tablenotetext{}{$\sigma_\mathrm{tel}$ -- telescope noise}
	\tablenotetext{}{$\sigma_\mathrm{\delta DM}$ -- DM estimation uncertainty: $\sigma_\mathrm{\delta DM}=\sqrt{\sigma_{\widehat{DM}}^2 + \sigma_{\delta t_C}^2 + 
	\sigma_{\mathrm{DM}(\nu)}^2}$}
	\tablenotetext{}{$\sigma_\mathrm{W}$ -- white noise errors}
	\tablenotetext{}{$\sigma_\mathrm{J}$ -- jitter noise}
	\tablenotetext{}{$A_\mathrm{red}$, $\gamma_\mathrm{red}$ -- red noise power spectrum amplitude and spectral index}
	\tablenotetext{}{$A_\mathrm{eff} / A_\mathrm{eff, tot}$ -- fraction of the total collecting area used in observation}
        \label{tab:psrlist}
    \end{splitdeluxetable}

Next, we consider two variants of our mock timing program: (1) an all-sky PTA consisting of only pulsars with $\sigma_{\rm TOA}<100$ ns (\texttt{all\_sky\_best}) and (2) a targeted campaign in an area of the sky consisting of $\sim 10$ pulsars within a $\sim 10^{\circ}$ radius, where the GW source is chosen to be located within that sky area (\texttt{targeted}). Since the antenna pattern functions increase with a closer (but not perfect) alignment between the source and the pulsar (see the definitions in \citealt{Ellis2012}), a PTA campaign which targets pulsars near an EM-selected source may be more advantageous for detecting its GW signal. We plot the pulsars in one realization of \texttt{all\_sky\_best} as large filled circles in the second panel of Figure \ref{fig:map}, and those in \texttt{targeted} as stars in the third panel. Their respective $\sigma_{\rm TOA}$ distributions are shown in the last two panels in Figure \ref{fig:sigma_hist}. The properties of a sample of simulated MSPs are shown in Table \ref{tab:psrlist}, and ten realizations of the mock population are available in machine-readable form online.

Then, for each pulsar sample in each realization, we construct mock PTA observations which have the baseline of $15$ years which is approximately the length of the current PTAs. The cadence of observations is chosen to be either monthly or weekly, to imitate the current ``monthly'' and ``high cadence'' campaigns of NANOGrav \citep{NG12pt5yrpsr}.

\begin{table}[ht]
\caption{SMBHB parameters}
\begin{center}
\begin{tabular}{ccccccc}
\hline \hline
Source & R.A.& decl. & $\mathcal{M}$ & $f_{\rm GW}$ & $d_{\rm L}$ & $\alpha$ \\
 & deg & deg & $\log M_{\odot}$ & $\log$ Hz & $\log$ Mpc & ns \\ 
 \hline
S1 & 133.7036 & 20.1085 & 9 & $-$8.28 & 3.2 & 3.4 \\ 
S2 & 300  & 30 & 9 & $-$8 & 2.5 & 13.9 \\ 
S3 & 250  & $-$10 & 9 & $-$8 & 2.5 & 13.9 \\ 
\hline \hline
\end{tabular}
\end{center}
\tablecomments{$\Psi$ = $\pi$/2, $\Phi_{0}$ = $\pi$/2, $i$ = $\pi$/4 for each source. S1 represents an OJ 287-like SMBHB, whereas S2 and S3 are mock SMBHBs.}
\label{tab:sources}
\end{table}

\begin{table}[ht]
\caption{Mock PTA campaigns}
\begin{center}
\begin{tabular}{cll}
\hline \hline
Source  & PTA & Campaign \\
\hline
S1 & \texttt{all\_sky} & 15 yr, monthly \\
 & \texttt{all\_sky\_best} & 15 yr, monthly \\
 & \texttt{all\_sky\_best} & 15 yr, weekly \\
\hline
S2 & \texttt{all\_sky} & 15 yr, monthly \\
 & \texttt{all\_sky\_best} & 15 yr, monthly \\
 & \texttt{all\_sky\_best} & 15 yr, weekly \\
 & \texttt{targeted} & 15 yr, monthly \\
 & \texttt{targeted} & 15 yr, weekly \\
 \hline
S3 & \texttt{all\_sky} & 15 yr, monthly \\
 & \texttt{all\_sky\_best} & 15 yr, monthly \\
 & \texttt{all\_sky\_best} & 15 yr, weekly \\
 \hline \hline
\end{tabular}
\end{center}
\tablecomments{Properties of the sources are listed in Table \ref{tab:sources}.}
\label{tab:campaigns}
\end{table}

\subsection{Parameters and GW signals of an SMBHB} \label{sec:signal}

We consider a binary system with component masses $m_{1}$ and $m_{2}$. The so-called chirp mass $\mathcal{M}$ is given by $\mathcal{M} \equiv (m_{1} m_{2})^{3/5}/(m_{1}+m_{2})^{1/5}$. We assume the system is in a circular orbit whose GW frequency is related to its orbital frequency $f_{\rm GW}=2 f_{\rm orb}$ and is often expressed as $\omega = \pi f_{\rm GW}$. The source is located at a luminosity distance $d_{\rm L}$; since in this work we will only consider nearby sources, we ignore the effect of redshift: $(1+z) \approx 1$. For simplicity, we only consider the ``Earth term'' in the timing residual and do not consider the ``pulsar term,''\footnote{Our justification for only considering the Earth term is mainly twofold: (1) our methodology (see \S\ref{sec:fisher}) is only concerned with the parameter measurement uncertainty and does not inform the possible bias in the posterior distribution introduced by dropping the pulsar term; (2) extracting source information from the pulsar term would require precise measurements of the pulsar distance which we do not assume in this work.} and hence only the combination $\mathcal{M}^{5/3}/d_{\rm L}$ can be constrained. It is often convenient to redefine the amplitude as $\alpha = \mathcal{M}^{5/3}/d_{\rm L}/\omega^{1/3}$.

The sky position of the source is written as polar and azimuthal angles $\phi$ and $\theta$ (corresponding to right ascension and declination in units of radians: $\phi=\rm R.A.$, $\theta = \pi/2 - \rm decl.$). We assume the frequency evolution of the binary over the course of $\sim 15$ years is negligible (i.e., a monochromatic signal). We further assume the BHs are non-spinning, since PTAs are largely insensitive to the effects of spin in an SMBHB system (see \citealt{Sesana2010}). 

Such a binary system is then described by the following parameters: $p = \{\alpha, \omega, \phi, \theta, i, \psi, \Phi_{0} \}$, where $i, \psi, \Phi_{0}$ are the inclination, GW polarization angle, and initial orbital phase, respectively. For simplicity, we adopt fixed values $\psi=\pi/2, \Phi_{0}=\pi/2$, and an intermediate inclination $i=\pi/4$ for each of the sources, since these are less astrophysical interesting parameters for our study. We then follow the usual definitions in \cite{Ellis2012} and \cite{NG11yrCW} to compute the Earth-term timing residual {\it Res} corresponding to each pulsar which has coordinate $\phi_{\rm psr}$ and $\theta_{\rm psr}$.

The first source (S1) which will be ``observed'' by our mock PTA is the circular and non-spinning analog of the well-known SMBHB candidate OJ 287 (e.g., \citealt{Sillanpaa1988,Lehto1996}; see e.g., \citealt{Dey2019,Valtonen2021} for a recent review). Adopting the binary parameters in \cite{Dey2018}, we estimate a GW amplitude $\alpha \approx 3$ ns. We note that while it is possible to compute a more accurate waveform for this complex binary candidate (see e.g., \citealt{Valtonen2021}), in this work we only seek to obtain estimates of its detection and parameter measurement prospects.

As can be seen in Figure \ref{fig:map}, S1 is located in an area of the sky where few mock pulsars are nearby (e.g. within a $\sim$ 10 deg radius), thereby making a future targeted pulsar timing campaign with DSA-2000 unlikely. Prompted by this, we consider a mock source (S2) which (1) has similar binary parameters but is located at the part of the sky where it may be observed by a targeted campaign and (2) represents a generic SMBHB candidate, whereas S1 is based on an actual candidate. We additionally consider an S3, who shares the same binary parameters as S2 but is at a less advantageous location for pulsar searching and timing (and hence the possibility of a targeted campaign). On the other hand, it is located outside the Galactic plane and suffers from less reddening and extinction effects. Thus, S3 represents sources which could be easily detected or observed by most EM observatories. The parameters of the three sources are listed in Table \ref{tab:sources}, where the S1 parameters are based on \cite{Dey2018}.

To summarize, we observe the three sources with two all-sky PTAs with either weekly or monthly cadence over 15 years. However, an \texttt{all\_sky} campaign at a weekly cadence is not under consideration here since the total observing time would exceed the 25\% available for pulsar timing. For S2, we additionally consider a targeted campaign at both cadences. Those mock programs are also summarized in Table \ref{tab:campaigns}.

\subsection{Estimating binary parameters using the Fisher matrix formalism} \label{sec:fisher}

The Fisher matrix is often used to assess the parameter estimation performance of an experiment in fields such as cosmology (e.g., \citealt{Albrecht2006}) and GW astrophysics (e.g., \citealt{Sesana2010,McGrath2021}). While it can lead to a less accurate prediction of the performance of the experiment (see e.g., \citealt{Vallisneri2008}) than the mock data analysis approach, it is significantly less computationally expensive and thus can be applied to a large number of experimental setups and realizations. In this section, we describe how we apply the Fisher matrix formalism to predict the prospects of mock PTAs as described in \S \ref{sec:pta} of measuring the parameters of the three SMBHBs as depicted in \S \ref{sec:signal}.

We begin by first computing the Fisher matrix for each pulsar (indexed by $a$):

\[
\mathcal{F}^{a}_{ij} = \sum_{b} \frac{1}{\sigma^{2}} \frac{\partial Res_{b}}{\partial p_{i}} \frac{\partial Res_{b}}{\partial p_{j}} \, ,
\]

\noindent where $i$ and $j$ are the indices for binary parameters, $b$ denotes each observation, $\sigma$ represents the $\sigma_{\rm TOA}$ of the $a$-th pulsar as discussed in \S \ref{sec:pta}\footnote{We have assumed white noise for the Fisher matrix formalism. Properly including pulsar red noise would require non-trivial modifications to the Fisher matrix, which is beyond the scope of this paper. However, for a discussion of the potential effect of red noise on detection and parameter estimation, see Section \ref{sec:hasasia}.}, and $Res$ is the Earth-term timing residual corresponding to the $a$-th pulsar as described in \S \ref{sec:signal}. Here we fix the values of $\phi$ and $\theta$ to mimic the search for a CW signal where the source has been pinpointed (which is a reasonable assumption if the candidate is identified electromagnetically), which means we do not compute the partial derivatives with respect to $\phi$ or $\theta$, and hence $i,j = 1...5$ which correspond to the parameters $\{\alpha, \omega, i, \psi, \Phi_{0} \}$.

We then compute the Fisher matrix for multiple pulsars at various locations \{$\phi_{\rm psr}$, $\theta_{\rm psr}$\} (i.e., a PTA) by summing the matrices:

\[
\mathcal{F} = \sum_{a} \mathcal{F}^{a} \, .
\]

Instead of directly inverting the matrix $\mathcal{F}$, we first ensure it is well-conditioned, by normalizing the matrix by the factor $\sqrt{\mathcal{F}_{ii} \mathcal{F}_{jj}}$ so that the diagonal elements are 1 and the off-diagonal elements are of order unity. We then use singular value decomposition to invert the normalized matrix: $\mathcal{C}_{N} = \mathcal{F}^{-1}_{N}$. We then divide $\mathcal{C}_{N}$ by the normalization factor to obtain the final covariance matrix $\mathcal{C}$, where $\mathcal{C}=\mathcal{F}^{-1}$. Finally, we check the accuracy of the inversion by multiplying the covariance matrix by the original Fisher matrix and confirm that its difference from the identity matrix is lower than an error threshold: max($|\mathbb{I}-\mathcal{F}\mathcal{C}|)<10^{-3}$. We can then obtain the measurement uncertainty $\sigma_{i}$ of parameter $p_{i}$ from the $i$-th diagonal element $\sigma^{2}_{i} = \mathcal{C}_{ii}$. In this work, we focus on the uncertainties of the two more astrophysically interesting parameters, $\alpha$ and $\omega$.


\begin{figure*}[ht]
\centering
\epsfig{file=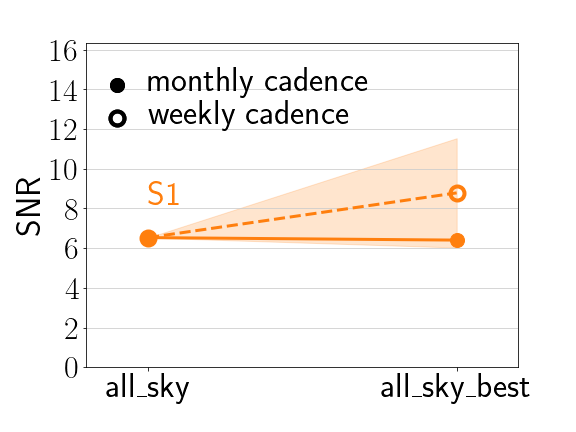,width=0.32\textwidth,clip=}
\epsfig{file=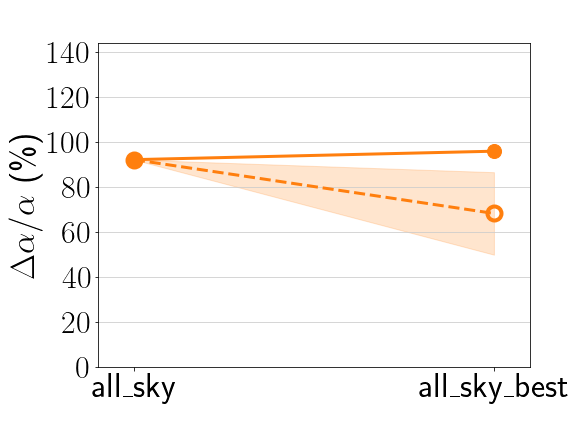,width=0.32\textwidth,clip=}
\epsfig{file=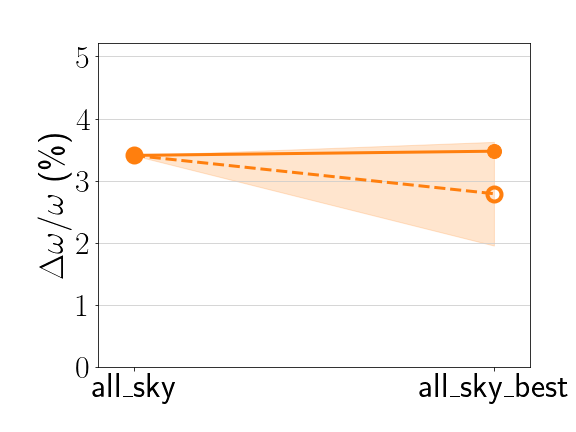,width=0.32\textwidth,clip=}
\caption{On the left panel, we plot the SNRs of S1 (based on the binary candidate OJ 287) as observed by two types of PTAs, \texttt{all\_sky} and \texttt{all\_sky\_best}, both with a 15 yr-long baseline. Recall that S1 is not observed by a \texttt{targeted} campaign (see text for details). 
On the right panels, we show the measurement uncertainties of $\alpha$ and $\omega$. The solid line compares the mean SNRs or parameter uncertainties for the two PTAs at the same, monthly observing cadence, whereas the dashed line compares \texttt{all\_sky} at a monthly cadence versus \texttt{all\_sky\_best} at a weekly cadence. The shaded bands represent 1-$\sigma$ uncertainties from the realizations, except for the monthly cadence case which does not have sufficient SNR$>5$ realizations to compute uncertainties.}
\label{fig:s1}
\end{figure*}


Finally, we estimate the total SNR by summing the contribution from each pulsar: $\rm SNR^{2}$ = $\sum_{a}$ SNR$_{a}^{2}$, where SNR$_{a}$ = $\sqrt{\sum_{b} (\frac{Res_{b}}{\sigma})^{2}}$. In this work, we focus our analysis in the strong signal regime (defined in this work as SNR$>5$) so that $\sigma_{i}$, instead of being an lower limit, approaches the actual measurement uncertainty. Therefore we only record the value of $\sigma_{i}$ if SNR is greater than 5 in a realization.

\section{Results} \label{sec:results}

\subsection{PTA observations of an OJ 287-like binary candidate} \label{sec:s1}

We first investigate whether S1 can be observed by a future all-sky PTA at a either weekly or monthly cadence, i.e. our simulated \texttt{all\_sky} and \texttt{all\_sky\_best}, by performing Fisher matrix analysis for each simulated PTA. At a monthly cadence, only 2/10 of the realizations of an \texttt{all\_sky} PTA exceed our SNR threshold of 5. Their SNRs slightly decrease in \texttt{all\_sky\_best}, which is expected because the pulsars in \texttt{all\_sky\_best} are a subset of the ones in \texttt{all\_sky} (solid line in Figure \ref{fig:s1}). If only the best pulsars (\texttt{all\_sky\_best}) are monitored at a weekly cadence, 9/10 of the realizations have SNR$>5$, and the mean value increases to $\sim 9$ (dashed line). 

We proceed to compare the measurement uncertainties between GW- and EM-based methods, where the GW-based uncertainties are computed using the Fisher matrix method in \S \ref{sec:fisher}, and the EM-based uncertainties are based on the reported uncertainties of $m_{1}$, $m_{2}$, and $P_{\rm orb}$ from  \cite{Dey2018}. The measurement uncertainty of the GW amplitude (quantified by $\Delta \alpha/\alpha$) is at the $\sim100\%$ level (middle panel), far exceeding the EM-based measurement uncertainty of $\sim0.2\%$ for this binary candidate. By contrast, $\omega$ can be constrained at the few percent level despite the modest SNR (right panel); it is however still greater than the EM-based uncertainty of $\sim0.1\%$.

Thus, we tentatively conclude that a future PTA with DSA-2000 which resembles our  \texttt{all\_sky} or \texttt{all\_sky\_best} could detect the GW signal from OJ 287 with $\sim15$ years of data at $>$monthly cadence. However, evidence for its binarity based only on GW observations will be modest, unless GW data (especially the more precise measurement of $f_{\rm GW}$) are interpreted alongside EM observations. Further, the stochastic GWB has a fiducial power-law spectral shape which increases in amplitude at low GW frequencies; therefore the GWB would have a significant effect on the detectability of sources emitting at (particularly) lower frequencies. See Section \ref{sec:hasasia} for a discussion of this effect on the three mock sources considered in this paper. 

It is worth noting that the \texttt{all\_sky} campaign only increases the SNR by $ <1$ compared to \texttt{all\_sky\_best}, despite monitoring $\sim6$ times the number of pulsars. This suggests that if the total number of observations is a limited resource, it may be best spent on the highest quality pulsars, either at the same cadence or a higher cadence. We explore this further in the following sections.

\begin{figure*}[ht]
\centering
\epsfig{file=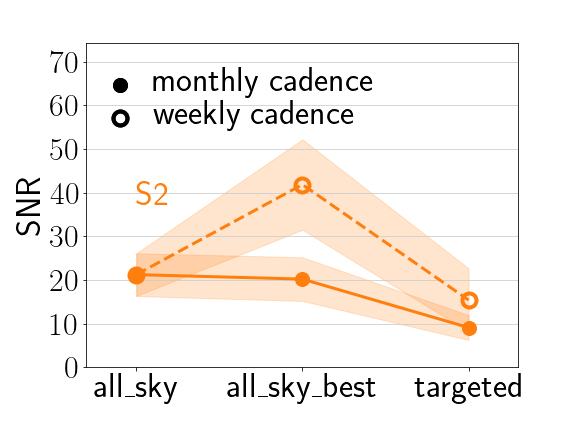,width=0.32\textwidth,clip=}
\epsfig{file=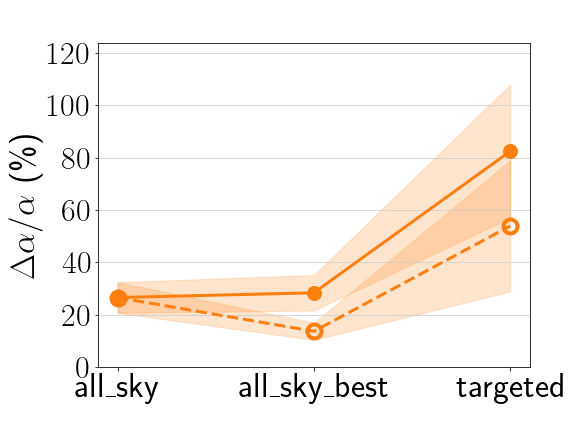,width=0.32\textwidth,clip=}
\epsfig{file=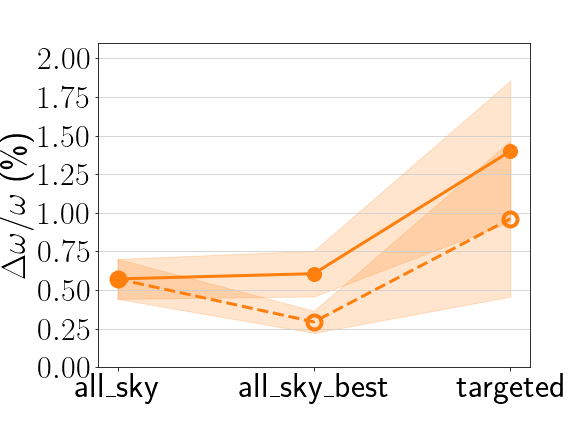,width=0.32\textwidth,clip=}
\caption{Same as Figure \ref{fig:s1}, but for a mock source S2. Additionally, results of the \texttt{targeted} campaign is included.}
\label{fig:s2}
\end{figure*}

\begin{figure*}[ht]
\centering
\epsfig{file=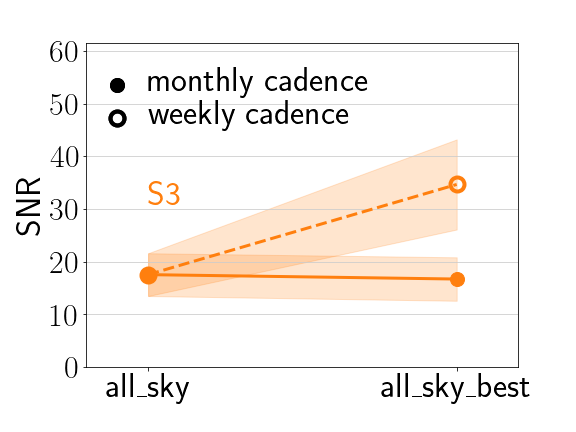,width=0.32\textwidth,clip=}
\epsfig{file=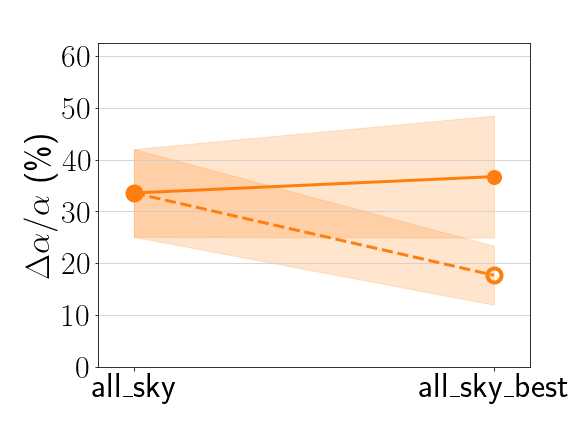,width=0.32\textwidth,clip=}
\epsfig{file=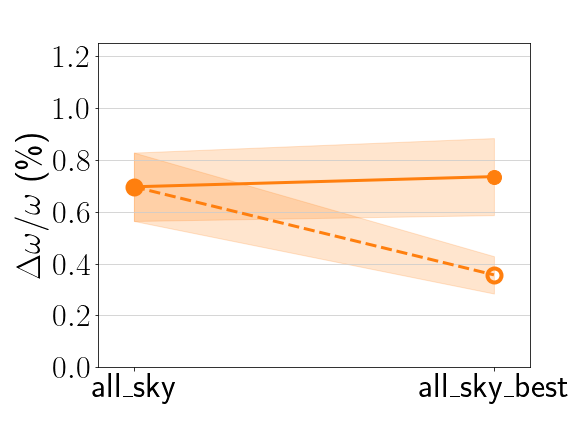,width=0.32\textwidth,clip=}
\caption{Same as Figure \ref{fig:s1}, but for S3. Recall that S3 is also not observed by a \texttt{targeted} campaign.}
\label{fig:s3}
\end{figure*}


\subsection{PTA observations of a mock binary candidate} \label{sec:s2}

In this section, we consider the detectability and parameter measurement uncertainties of the mock source S2, which has a larger GW amplitude ($\alpha \approx 14$ ns) and is located in a part of the sky where more pulsars may be found and monitored. We first compute the expected results of a monthly, 15 yr-long campaign. As we show in the left panel of Figure \ref{fig:s2}, the highest SNR is reached in the \texttt{all\_sky} campaign; it decreases in \texttt{all\_sky\_best} and decreases further in \texttt{targeted}. This can be understood by the fact that \texttt{targeted} has comparable pulsar timing noise as \texttt{all\_sky}, but $\sim10$ times fewer pulsars in the array. Similarly, \texttt{all\_sky\_best} has much fewer pulsars than \texttt{all\_sky} and hence a lower SNR, despite a higher median pulsar timing quality.

As we also show in the last two panels of Figure \ref{fig:s2}, \texttt{targeted} results in poorer constraints on $\alpha$ and $\omega$, consistent with its lowest SNR. Nonetheless, these GW-based parameter measurements are better than the constraints obtained from typical EM observations for the following reasons. (Recall that S2 represents a generic EM binary candidate and therefore its parameters are assumed to be measured from standard methods.) First, an observed variability period caused by modulated accretion in the binary may not directly correspond to its intrinsic orbital period, since this relationship may be mass ratio-dependent and the observed period may be at a few times the orbital period (see e.g., \citealt{D'Orazio2013,Farris2014,Noble2021}); therefore, an EM-based frequency measurement would have an (underestimated) uncertainty of $100\%$, which is $\sim 2$ orders of magnitude larger than the GW-basement measurement (right panel in Figure \ref{fig:s2}). Second, standard BH mass measurements suffer from a systematic uncertainty of $\sim 0.3$ dex (e.g., \citealt{Shen2013}); combined with the aforementioned uncertainty on $\omega$, this translates to a $\Delta\alpha/\alpha$ of approximately 120\%, which is a few times higher than the GW-based measurement (middle panel in Figure \ref{fig:s2}). These precise GW parameter measurements can allow meaningful comparisons with EM-based measurements as a robust test of the binary model, breaking the degeneracies arising from EM-only observations, and testing the predictions of SMBHB theory.

It is however important to note that \texttt{all\_sky} boasts the highest SNR simply due to the fact that it is timing $\sim 6$ times more pulsars (than \texttt{all\_sky\_best}) and that the contribution of bottom 85\% of the pulsars to the SNR {\it combined} only increases the SNR by $\sim 1$. Therefore it would be interesting to investigate whether \texttt{all\_sky\_best} or \texttt{targeted} is a more efficient approach to achieve a similar scientific outcome.

To do so, we compare the results of the high-cadence, weekly \texttt{all\_sky\_best} and \texttt{targeted} campaigns with the (monthly) \texttt{all\_sky} campaign. As shown in Figure \ref{fig:s2}, we predict that \texttt{all\_sky\_best} results in the highest SNR and best parameter measurements; this is despite the fact that \texttt{all\_sky\_best} costs $\sim$ 40\% less total telescope time than \texttt{all\_sky}. Likewise, a high-cadence \texttt{targeted} campaign requires only $\sim$a third of the time as \texttt{all\_sky}, but achieves a similar SNR. Furthermore, both high-cadence campaigns can measure $\alpha$ and $\omega$ at precise levels which are better than the typical EM measurements as discussed previously. Therefore, we conclude that \texttt{all\_sky\_best} and \texttt{targeted} are both possible, if not preferable, alternatives to \texttt{all\_sky}.

\subsection{PTA observations of a second mock binary candidate} \label{sec:s3}

Finally, we examine the detection and parameter estimation prospects of S3. As we show in Figure \ref{fig:s3}, the expected outcome of \texttt{all\_sky} and \texttt{all\_sky\_best} follows the same trend; namely, \texttt{all\_sky} results in a higher SNR for the same cadence, but the high-cadence \texttt{all\_sky\_best} campaign outperforms \texttt{all\_sky} at a lower cost of total telescope time. Further, despite the slightly lower SNR than S2 (consistent with the difference in sky location), the parameters of S3 can still be measured to moderate to high precision.

As discussed earlier, the sky location of S3 may be more representative of that of a likely EM candidate: it is located where it is more likely to be found by EM observatories, including all-sky surveys; on the other hand, it is not at such a fortuitous location that it happens to be in close proximity to a large number of (high quality) MSPs. Fortunately, its detection prospects are nonetheless good in both all-sky campaigns and are only mildly impacted by its ``inferior'' location. This suggests that, contrary to conventional wisdom, dedicated timing campaigns targeting pulsars near likely sources may not be necessary, efficient, or possible for CW observations.


\subsection{Effects of the GWB and pulsar red noise on the detectability of S1--S3}\label{sec:hasasia}

Recently, a red noise process which has a common spectrum across pulsars has been observed in the regional PTAs and the combined International Pulsar Timing Array dataset \citep{NG12p5GWB,EPTACRN,PPTACRN,IPTADR2}. While the PTAs have not confirmed this common red noise process as the GWB, it already needs to be accounted for in more recent CW searches (e.g. NANOGrav Collaboration, in prep.) and will have an effect on PTA's sensitivity to CWs, especially at lower frequencies. Further, some level of red noise is present in many, if not most, MSPs, which would further compromise the observability of CW sources.

While our Fisher matrix methodology does not handle red noise, in this section, we seek to understand the potential effect of the GWB and pulsar red noise on the detectability of CW sources using simulated PTA sensitivity curves and speculate their implications for binary parameter estimation.

The GWB has a power law power spectrum in the form:

\begin{equation*}
h_{c} (f) = A_{\rm GWB}\Big(\frac{f}{\rm yr^{-1}}\Big)^{\alpha} \, ,
\end{equation*}

\noindent where we adopt the median value of the amplitude from the NANOGrav 12.5 yr analysis: $A_{\rm GWB} = 1.92 \times 10^{-15}$ \citep{NG12p5GWB}, and $\alpha = -2/3$ \citep{Phinney2001}.

The power spectrum of the pulsar red noise is given by (e.g., \citealt{NG12p5yrdata}):

\begin{equation*}
P(f) = A_{\rm RN}^{2} \Big(\frac{f}{\rm yr^{-1}}\Big)^{\gamma_{\rm RN}} \, ,
\end{equation*}

\noindent where $A_{\rm RN}$ and $\gamma_{\rm RN}$ are the amplitude and the index, respectively.
We simulate $A_{\rm RN}$ and $\gamma_{\rm RN}$ for each mock pulsar using the model developed by
\citet{Shannon2010}. They determined that the natural log of the measured,
red timing noise after a second-order polynomial fit is normally distributed as
$\ln (\sigma_\mathrm{TN,2})\sim \mathcal{N}\left(\ln (\hat{\sigma}_\mathrm{TN,2}), \delta\right)$.
$\hat{\sigma}_\mathrm{TN,2}$ is the post-fit red noise scaled to the spin period $P$ and period derivative
$\dot{P}$ given by
\begin{equation}\label{eq:sigredhat}
    \hat{\sigma}_\mathrm{TN,2} = 
							10^{15}C_2 P^{-\alpha}\left(\frac{\dot{P}}{P^2}\right)^\beta T_\mathrm{yr}^\gamma\mathrm{\ ns},
\end{equation}
where $C_2$, $\alpha$, and $\beta$ are estimated over the pulsar population and $T_\mathrm{yr}$ is the timescale
of the residuals in years. We assume that spin period derivatives are uncorrelated with spin period and are distributed log-normally with  $\langle \log_{10}\dot{P} \rm \rangle= -19.9$ and $\sigma_{\log_{10}\dot{P}} = 0.5$. We adopt the updated values for $C_2$, $\alpha$, $\beta$, $\delta$, and $\gamma$, estimated for a larger sample of MSP red noise measurements, in Table 4 of \citet{Lam2017}, row ``$\mathrm{MSP_{10,PPTA}} + \mathrm{NANO}$." The red noise index, $\gamma_{\rm RN} = -(2\gamma + 1)$, is assumed to be -5.6 for all MSPs. By combining Equation \ref{eq:sigredhat} and Equation 15 of \cite{Lam2017} for the measured red noise in terms of $A_{\rm RN}$ and $\gamma_{\rm RN}$, we can directly draw $A_{\rm RN}$ for each model pulsar where

\begin{figure*}[ht]
\centering
\epsfig{file=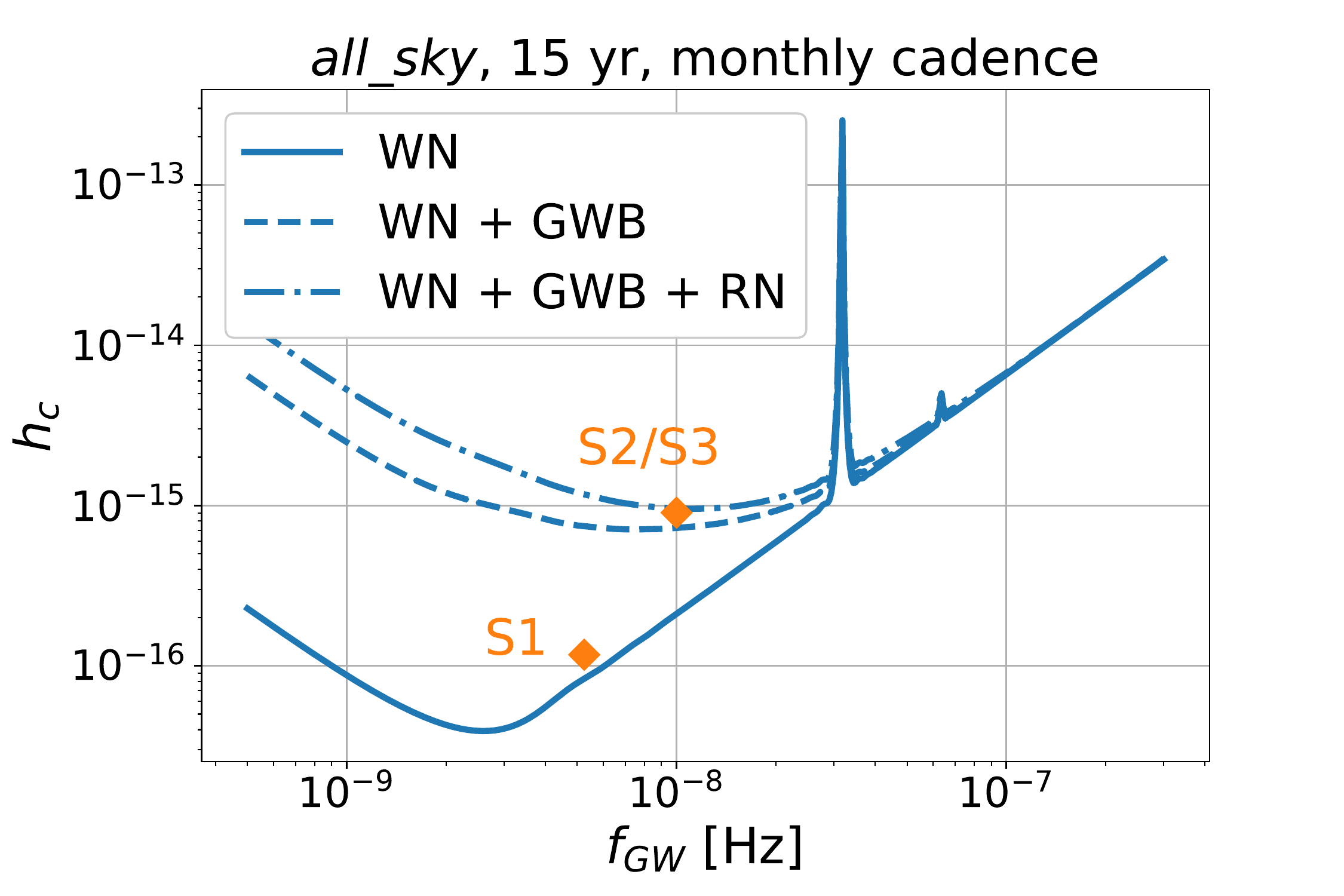,width=0.5\textwidth,clip=}
\caption{We show the expected decrease in sensitivity to CWs by the PTA in the presence of (1) only pulsar white noise (solid curve), (2) pulsar white noise and the GWB (dashed curve), and (3) pulsar white noise, GWB, and pulsar red noise (dash-dotted curve). We superimpose the GW strains and frequencies of the three mock sources (orange diamonds) for visual purposes. 
}
\label{fig:hasasia}
\end{figure*}

\begin{equation*}
    f({A}_\mathrm{RN}) = \frac{1}{\sqrt{2\pi\delta^2}}\
                                                        \exp\left[-\left(\frac{\left(\ln(\hat{A}_\mathrm{RN}) - \ln(A_\mathrm{RN})\right)^2}{2\delta^2}\right)\right]
\end{equation*}
and
\begin{equation*}
               \hat{A}_\mathrm{RN}  = (-(1 + \gamma_\mathrm{RN}))^{1/2}10^{15} C_2
               P^{-\alpha}\left(\frac{\dot{P}}{P^2}\right)^\beta\mathrm{\mu s\ yr^{1/2}}.
\end{equation*}

To simulate the PTA sensitivity to CWs in the presence of these noise terms, we use the software package \texttt{hasasia}\footnote{https://github.com/Hazboun6/hasasia} \citep{Hazboun2019}. We first compute the sensitivity curve by using the sky positions of simulated \texttt{all\_sky} pulsars and only including white timing noise (see \S\ref{sec:pta}). Following the previous sections, we assume a 15-yr PTA with a monthly cadence. The resulting sensitivity is shown as the solid curve in Figure \ref{fig:hasasia}. For a monochromatic source (which we assume throughout the paper) emitting at $f_{0}$ with a strain amplitude $h_{0}$, the expected SNR is $h_{0}\sqrt{T_{\rm obs}/S_{\rm eff}(f_{0})}$, where $S_{\rm eff}$ is the effective strain-noise power spectral density for the PTA and is related to the sensitivity curve: $h_{c}(f) = \sqrt{f S_{\rm eff}(f)}$ \citep{Hazboun2019}. Therefore, the resulting SNRs for S1 and S2/S3 are 2.2 and 9.3, respectively, consistent with our results in \S\ref{sec:results} that the sources are detectable at the intermediate to high SNR level at monthly cadence.\footnote{Note however that the sensitivity curve is averaged over the sky and is not directly applicable to targeted searches. Therefore, the sensitivity curves in Figure \ref{fig:hasasia} are meant for comparison with each other for the purpose of understanding the effect of red noise.} 

Next, we recompute the sensitivity in the presence of the GWB. The result is shown as the dashed curve. As expected, the GWB severely decreases the PTA sensitivity to CWs, especially at frequencies lower than $\sim 2\times 10^{-8}$ Hz where it decreases by up to $\sim$ an order of magnitude. Consequently, S1 would not be detectable above the GWB (SNR=0.2), while S2 and S3 may still be detectable, albeit with a lower SNR (2.7).

Finally, we include the pulsar red noise and show the resulting sensitivity curve as the dash-dotted curve. The effect is less pronounced, with the sensitivity decreasing by less than a factor of a few at all frequencies, resulting in an SNR = 0.2 (2.1) for S1 (S2/S3).

We further anticipate that PTA's ability to estimate binary parameters would also decrease in the presence of red noise. This is based on the fact that at a given frequency, source detectability and parameter measurability closely track each other (see Figures \ref{fig:s1} -- \ref{fig:s3}), i.e., a lower SNR corresponds to a larger measurement uncertainty.

While our Fisher matrix method cannot handle red noise and hence our conclusions in this work regarding source detectability and observing strategies are drawn assuming only white noise (i.e., solid curve in Figure \ref{fig:hasasia}), future work would be able to make more accurate predictions of the detectability of CW sources by taking into account the additional noise, especially the GWB. In practice, the (relatively small) effect of pulsar red noise could be mitigated by only including pulsars with low red noise levels in the PTA, while the effect of the astrophysical GWB cannot removed.


\section{Summary and Conclusions} \label{sec:conclude}

In this work, we have applied the Fisher matrix formalism to predict the SMBHB parameter estimation prospects of a future PTA with DSA-2000. We focus on the case where the sky location of the source is known via the prior identification of a possible EM counterpart, motivated by (1) the boosted detection SNR and parameter measurements by searching for the GW signal at a fixed sky location, and (2) the possibility of studying SMBHBs in a true multi-messenger fashion. Compared to previous work, we have taken a more realistic approach, taking into consideration factors such as practical and observational constraints of a PTA, the Galactic pulsar population, the number and timing precision of MSPs detectable by a telescope, and their spatial distribution. 

As a case study, we first examined the detection prospects of the well-known SMBHB candidate OJ 287 (using S1 as a surrogate). We estimate that a future PTA with DSA-2000 may be able to detect the source with 15 years of data (assuming only pulsar white noise). While the measurement uncertainty of the GW amplitude is large, which translates to a poor constraint on the system mass, the PTA constraint on the GW frequency is at a level which is more comparable to the EM equivalent, permitting the possibility of robustly testing its binary model through a joint interpretation of GW and EM data.

We then generalized this approach to a mock source, and our main results can be summarized as follows:

\begin{enumerate}
\item We have considered three types of PTAs, an all-sky PTA which observes all pulsars in our mock pulsar sample (\texttt{all\_sky}), an all-sky PTA which only observes the best quality pulsars (\texttt{all\_sky\_best}), and a PTA which targets pulsars near an EM-selected binary candidate (\texttt{targeted}). At equal observing cadence, \texttt{all\_sky} naturally results in the highest SNR and best parameter measurements. However, even with less total observing time, the high-cadence \texttt{all\_sky\_best} outperforms the other ones.
\item We have examined the detection and parameter estimation prospects of the mock source placed at different sky locations. If it is at a fortuitous location where a higher-cadence targeted campaign is possible (i.e., S2), the \texttt{targeted} campaign yields results similar to a lower-cadence \texttt{all\_sky} campaign. More realistic is a sky location where \texttt{targeted} is not feasible due to the dearth of pulsars near the source (i.e., S3); this scenario may be more likely for EM-selected binary candidates. Fortunately, at such a location, its detection prospects with \texttt{all\_sky} or \texttt{all\_sky\_best} may only decrease slightly compared to S2.
\item For a more typical EM candidate (represented by S2 and S3), the GW amplitude and frequency parameters may be constrained at a level which is better than standard EM measurements. Particularly, GW frequency can be easily constrained within a few percent; this level of precision is consistent with similar estimates from previous work using both mock data analysis (e.g., \citealt{Liu2021}) and Fisher matrix-based approaches (e.g., \citealt{Sesana2010}).
\end{enumerate}

Those results suggest the following strategies for a PTA to achieve the best science outcome at the lowest cost:

\begin{enumerate}
\item Ideally, an all-sky PTA timing all pulsars detectable by pulsar surveys at a high cadence yields the best outcome. In reality, such a PTA costs more time than what is available on a single telescope or array. A PTA which only times the best pulsars can achieve similar results at a lower cost of total telescope time, even if observing at a higher cadence. 
\item Applying the principle of choosing quality over quantity, an upgraded and downsized version of the present-day PTAs would be capable of detecting EM-selected SMBHBs with a variety of source parameters and measuring their parameters at precision levels which are better than conventional EM measurements. Such a PTA can operate as a true GW observatory of SMBHBs and complement EM searches and observations with profound implications for the multi-messenger studies of SMBHBs.
\item A dedicated PTA which targets a small number of MSPs near the EM-selected candidate may not be feasible, since the sky locations where MSPs can be found and where EM binary candidates may be observed likely do not coincide. Fortunately, this may not be necessary either, because an all-sky PTA consisting of a small number of high quality pulsars is capable of observing a source at {\it any} location with good results.
\end{enumerate}

We end with a caveat about our preference for \texttt{all\_sky\_best} over \texttt{all\_sky}. We have assumed known sky locations and therefore have not considered the PTA measurement uncertainties of $\phi$ and $\theta$ in this work. However, it has been known that sky localization improves with the number pulsars in the PTA \citep{Sesana2010}. Therefore, it cannot be determined from our study whether \texttt{all\_sky\_best} remains the best strategy for an unguided search over the whole sky without an EM counterpart beforehand. It would therefore be interesting to investigate the source localization capability of \texttt{all\_sky\_best} (with its more realistic pulsar spatial distribution and $\sigma$ distribution) and to reexamine its overall single source detection capability. 


\begin{acknowledgements}

This work is supported by the NANOGrav National Science Foundation Physics Frontiers Center award no.~2020265. The material is based upon work supported by NASA under award number 80GSFC21M0002. T.L. thanks Jolien Creighton, Jeff Hazboun, and David Kaplan for helpful suggestions. We thank the referee for useful comments that improved the paper.

\end{acknowledgements}

\software{\texttt{astropy} \citep{Astropy},  
	\texttt{FrequencyOptimizer} \citep{lam18_optimum_freq},
          \texttt{hasasia} \citep{Hazboun2019}, 
          \texttt{matplotlib} \citep{Matplotlib},
          \texttt{PsrPopPy} \citep{Bates2014}
          }
          


\bibliographystyle{aasjournal}

\end{document}